\documentclass[useAMS,usenatbib]{mn2e}

\usepackage{graphicx}
\usepackage{rotating}
\usepackage{caption}
\usepackage{subcaption}

\usepackage{color}
\usepackage{amsmath}
\usepackage{multirow}

\newlength{\colwidth}
\newcommand{\Msol}{\, \rm M_{\odot}}
\newcommand{\hMsol}{\, h^{-1}{\rm M_{\odot}}}
\newcommand{\hMpc}{\, h^{-1}{\rm Mpc}}

\newcommand{\hpc}{\, h^{-1}{\rm pc}}

\newcommand{\HI}{{H{\sc i}\,\,}}
\newcommand{\HII}{{H{\sc ii}\,\,}}

\newcommand{\zr}{z_{\rm r}}

\def\aj{AJ}					% Astronomical Journal
\def\apj{ApJ}					% Astrophysical Journal
\def\apjl{ApJL}					% Astrophysical Journal, Letters
\def\apjs{ApJS}					% Astrophysical Journal, Supplement
\def\aap{A\&A}					% Astronomy and Astrophysics
\def\mnras{MNRAS}				% Monthly Notices of the RAS
\def\pasp{PASP}					% Publications of the ASP
\def\nat{Nature}				% Nature
\begin{document}

\title{Low-mass galaxy formation and the ionizing photon budget during reionization}

\author[A. R. Duffy]
{Alan R. Duffy$^{1,2}$, J. Stuart B. Wyithe$^{1}$, Simon J. Mutch$^{1}$  and Gregory B. Poole$^{1}$  \\
$^1$School of Physics, University of Melbourne, Parkville, VIC 3010, Australia, \\
$^2$ARC Centre of Excellence for All-sky Astrophysics (CAASTRO)}

\date{}

\maketitle

\label{firstpage}

\begin{abstract}
We use high-resolution simulations of cosmological volumes to model galaxy formation at high-redshift, with the goal
of studying the photon budget for reionization. We demonstrate that galaxy formation models that include a strong, thermally 
coupled supernovae scheme reproduce current observations of star formation rates and specific star formation rates, both
during and after the reionization era.
These models produce enough UV photons to sustain reionization at $z{\la}8$ ($z{\la}6$) through a significant 
population of faint, unobserved, galaxies for an assumed escape fraction of 20\% (5\%).
This predicted population is consistent with extrapolation of the faint end of observed UV luminosity functions.
We find that heating from a global UV/X-ray background after reionization causes a dip in the total global star formation rate density
in galaxies below the current observational threshold. Finally, while the currently observed specific star
formation rates are incapable of differentiating between supernovae feedback models, sufficiently deep observations 
will be able to use this diagnostic in the future to investigate galaxy formation at high-redshift.
\end{abstract}


\section{Introduction}
\label{Introduction}

Understanding the Epoch of reionization, in which a primarily neutral Universe underwent a phase transition
to an almost completely ionized intergalactic medium (IGM) is a key driver in numerous theoretical and observational projects. 
This transition is a challenge to study observationally due to the faintness of the sources responsible for reionization, and 
because they are obscured by the neutral medium at optical wavelengths. It is also a challenge
to theoretically model this epoch through supercomputer simulations 
due to the enormous dynamic range between the small physical scales of the sources and the large volumes they affect. 
It is clear that the process of reionization is all but complete by redshift $z{\sim}6$ due to the presence of the~\citet{Gunn:65} trough
in quasar spectra at higher redshifts~\citep[e.g.][]{Fan:06}. 
Observations of cosmic microwave background (CMB) photons~\citep{Hinshaw:13} are consistent with an instantaneous 
reionization model that occurs at $z{=}10.3\pm 1.1$, implying that a more realistic `extended' reionization was well underway by this time.

\begin{table*}
\begin{center}
\caption{A list of the simulated galaxy formation physics schemes utilized in this study. From left to right, the 
columns list the simulation name used in this study, the name as defined by~\citet{Schaye:10} where appropriate (else `--' used to indicate a new scheme),
and a brief the description of the physics modelled. Each of these simulations has been run twice to test the effects of reionization on early galaxy formation. When we
add the label \emph{LateRe} (\emph{EarlyRe}) to the simulations names below we have modelled an instantaneous reionization
at $\zr{=}6.5$ $(9)$ with a~\citet{HaardtMadau:01} UV/X-ray background thereafter (see Section~\ref{sec:radeffect} for more details).}
\label{tab:physics}
\begin{tabular}{lll}
\hline
Simulation & {\sc OWLS} name & Brief description\\
\hline
\emph{PrimC\_NoSNe} & \emph{NOSN\_NOZCOOL} & No energy feedback from SNe and  cooling assumes primordial abundances\\
\emph{PrimC\_WSNe\_Kinetic} & \emph{NOZCOOL}  & Cooling as \emph{PrimC\_NoSNe} but with fixed mass loading of `kinetic' winds from SNe \\
\emph{ZC\_WSNe\_Kinetic} & \emph{REF} & SNe feedback as \emph{PrimC\_WSNe\_Kinetic} but now cooling rates include metal-line emission \\
\emph{ZC\_SSNe\_Kinetic} & \emph{--} & Strong SNe feedback as \emph{ZC\_WSNe\_Kinetic} but using 100\% of available SNe energy \\
\emph{ZC\_SSNe\_Thermal} & \emph{WTHERMAL} & Strong SNe feedback as \emph{ZC\_SSNe\_Kinetic} but feedback `thermally' coupled \\
\hline
\end{tabular}
\end{center}
\end{table*}

Recent ultra deep observations using the {\it Hubble Space Telescope} ({\it HST}) have begun constraining the galaxy UV luminosity function 
during the heart of this reionization era at $z{\sim}7 - 8$~\citep[e.g.][]{Bouwens:11a,Schenker:13, McLure:13}.~\citet{Robertson:13} extrapolated
the observed luminosity functions four magnitudes deeper than the {\it Hubble Space Telescope Ultra Deep Field} (HUDF) limits~\citep{Ellis:13,Koekemoer:13}
and found that these faint objects could sustain reionization given reasonable assumptions for the number of ionizing photons 
from the galaxies and clumping factor of the IGM. In this work, we demonstrate that this extrapolation is supported by our 
simulated galaxy populations, which exhibit a rising luminosity function below the observational magnitude limits 
that is sufficient to reionize the universe by $z{\la}8$.

As well as constraining luminosity functions, observations are also beginning to probe the nature of galaxies at high-redshifts. 
In particular, galaxies at higher redshifts have a higher specific star formation rate~\citep[sSFR; e.g.][]{Stark:13}. 
This had been predicted theoretically~\citep[e.g.][]{Bouche:10,Dave:11a} but was in disagreement with observations at the 
time which found a rapid rise in the sSFR from today to $z{\sim}2$ and then little or no evolution at higher redshift~\citep[e.g.][]{Daddi:07,Noeske:07,Stark:09,Gonzalez:10}.
We discuss this result in the light of different models of supernovae (SNe) feedback in this work. We find that current surveys have little discriminating
power, but that deeper observations can strongly discriminate between the feedback models, and determine the approximate redshift of reionization, through a combination of SFR and sSFR measurements.

The rest of this paper is organised as follows. In Section~\ref{sec:simulations} we introduce the new simulation suite {\sc Smaug} and the
range of galaxy formation physics we have examined. We then discuss the physics driving the SFR function (Section~\ref{sec:sfr}) 
and the sSFR (Section~\ref{sec:ssfr}). We show in Section~\ref{sec:sfrd} that our simulations support the 
paradigm of a `Photon-Starved'  Epoch of reionization~\citep{Bolton:07}, in which the number of ionizing photons is only {\it just} sufficient to sustain global reionization. 
In Section~\ref{sec:faintgal}, we demonstrate that there exists a faint galaxy population below observational limits that is consistent with extrapolating recent luminosity functions.
We conclude in Section~\ref{sec:conclusion}.

\section{Simulation Details}
\label{sec:simulations}
We have created a suite of high-resolution, cosmological volume hydrodynamical simulations,
named {\sc Smaug}, as part of the Dark Ages reionization and Galaxy Formation Simulations project ({\sc DRAGONS}). 
The {\sc Smaug} series of simulations all share the same initial conditions, i.e. are 
identical simulation volumes, but are resimulated using a number of different physics models (selected as a subset from the
OverWhelmingly Large Simulations series {\sc OWLS};~\citealt{Schaye:10}) which allows robust 
testing of the relative impact that each galaxy physics component has on observables such as SFRs for example.
These physics models have seen a number of successes in reproducing observed galaxy properties at low, post-reionization, redshifts.
For example, the cosmic star formation history~\citep[SFH; e.g.][]{Madau:96} was 
shown to be reproduced in~\citet{Schaye:10}, as well as the high-mass end of the \HI mass function in~\citet{Duffy:12a}.

The {\sc Smaug} simulations were run with an updated version of the publicly available {\sc Gadget-2} $N$-body/hydrodynamics 
code~\citep{Springel2005b} with new modules for radiative cooling~\citep{Wiersma:09a}, star formation~\citep{Schaye:08},
stellar evolution and mass-loss~\citep{Wiersma:09b} and galactic winds driven by SNe~\citep{DallaVecchia:08,DallaVecchia:12}. 
Each simulation employed $N^3$ dark matter (DM) particles, and $N^3$ gas particles, where $N{=}512$ for the production runs 
presented here (lower resolution test cases with $N{=}128$ and $256$ were also run for extensive resolution testing, see Duffy et al. in preparation)
within cubic volumes of comoving length $L{=}10\hMpc$. The Plummer-equivalent comoving softening length is $200 \hpc$ and the DM (gas/star) particle mass is $4.7\, (0.9) \times 10^{5} \hMsol$ respectively, for the $N{=}512$ production runs.
For all cases, we used grid-based cosmological initial conditions  generated with {\sc GRAFIC}~\citep{Bertschinger:01} at 
$z{=}199$ using the Zeldovich approximation and a transfer function from {\sc COSMICS}~\citep{Bertschinger:95}. 

We use the {\it Wilkinson Microwave Anisotropy Probe} 7-year results~\citep{Komatsu:11}, 
henceforth known as WMAP7, to set the cosmological parameters 
$[\Omega_{\rm m},$ $\Omega_{\rm b},$ $\Omega_{\Lambda},$ $h,$ $\sigma_{8},$ $n_{\rm s}]$
to [0.275, 0.0458, 0.725, 0.702, 0.816, 0.968] and 
$f^{\rm univ}_{\rm b}{=}0.167$. 

\subsection{Galaxy Formation Physics}
The naming convention for the simulations follows that of~\citet{Duffy:10}, with \emph{PrimC} indicating that only 
cooling from primordial elements (i.e. hydrogen and helium) is included. 
With simulations named \emph{ZC} we have included metal line emission cooling from hydrogen, helium and nine elements:
carbon, nitrogen, oxygen, neon, magnesium, silicon, sulphur, calcium and iron (calculated from the photoionization package {\sc CLOUDY}, last described in~\citealt{Cloudy}).

For the high-redshift simulations presented in this work, we have focused on testing the role of SNe feedback on early galaxy formation including the
effects of metals realized by these events enhancing the cooling rates of gas. We run a number of these models, from the same initial condition, but systematically 
increasing the complexity of the models to test the {\it relative} impact of the galaxy physics in observables such as SFR. 
Our first simulation is the most simplistic model possible, that of primordial gas cooling (\emph{PrimC})
that can then form stellar populations without a resultant phase of SNe feedback (\emph{NoSNe}). 
This first case is referred to as \emph{PrimC\_NoSNe} and results in runaway star formation as expected.

We then consider the cases where the SNe is able to couple a fraction of the available energy to nearby gas. 
The label \emph{WSNe} refers to weak feedback SNe (i.e. 40\% of the available SNe energy is coupled) 
and \emph{SSNe} refers to strong (i.e. 100\% available SNe energy coupled). 
There are two main techniques to couple this energy. The first uses SNe energy to heat nearby gas particles and 
to allow an outflow to build from the resultant overpressurized gas~\citep{DallaVecchia:12}. We term this \emph{Thermal} feedback. 
The alternative is to skip straight to the `snowplough' phase of the SNe feedback 
and directly `kick' gas particles~\citep{DallaVecchia:08} which we call \emph{Kinetic} feedback. 

Due to radiative losses, the \emph{Kinetic} feedback will have only a fraction of the available SNe energy $f_{\rm SNe}$ coupled into driving a wind (i.e. `kicking' the gas particles in this case),
while the \emph{Thermal} scheme implicitly models these losses~\citep{DallaVecchia:12} and hence is modelled using a higher fraction of the SNe energy ($f_{\rm SNe}{=}0.4,\,1.0$ respectively). 
To test if differences between the \emph{Kinetic} and \emph{Thermal} models was a result of numerical modelling, or simply the different fraction of SNe energy used, we also ran a 
maximal `strong' \emph{Kinetic} scheme with $f_{\rm SNe}{=}1.0$, in Duffy et al. (in preparation). We found that this `strong' \emph{Kinetic} scheme was in excellent agreement with the \emph{Thermal} model, 
indicating our resolution is sufficiently high that the method of coupling to the gas is not as important as the amount of energy released.
Therefore, although for completeness we have included \emph{ZC\_SSNe\_Kinetic} in Table~\ref{tab:physics} it is effectively indistinguishable 
from \emph{ZC\_SSNe\_Thermal} in the plots shown in this work and hence is not included for clarity.

The assumed initial mass function (IMF) for star formation in the {\sc Smaug} simulation series is from~\citet{Chabrier:03}.
When comparing with observational works which use the~\citet{Salpeter:55} IMF we reduce their SFR and stellar masses by a factor $1.65$. 
This reduction has been shown by~\citet{Haas:13a} to adequately address the conversion from UV photon counts from massive stars using the two IMFs.

The stellar feedback schemes we consider in this work are in good agreement with observed stellar mass functions and sSFRs at 
$z{=}0-2$~\citep{Haas:13a} at higher mass scales (their fig 4). These are therefore a compelling subset of simulations to consider at higher redshift.
We refer the reader to~\citet{Duffy:10} and~\citet{Schaye:10} for further details of the simulations. We provide a summary of the naming convention in 
Table~\ref{tab:physics}. In a future work, we will examine the impact that changing star formation prescriptions have on high-redshift galaxy formation. 
In this work, we have focused predominantly on the role that feedback plays for these objects.

\begin{figure*}
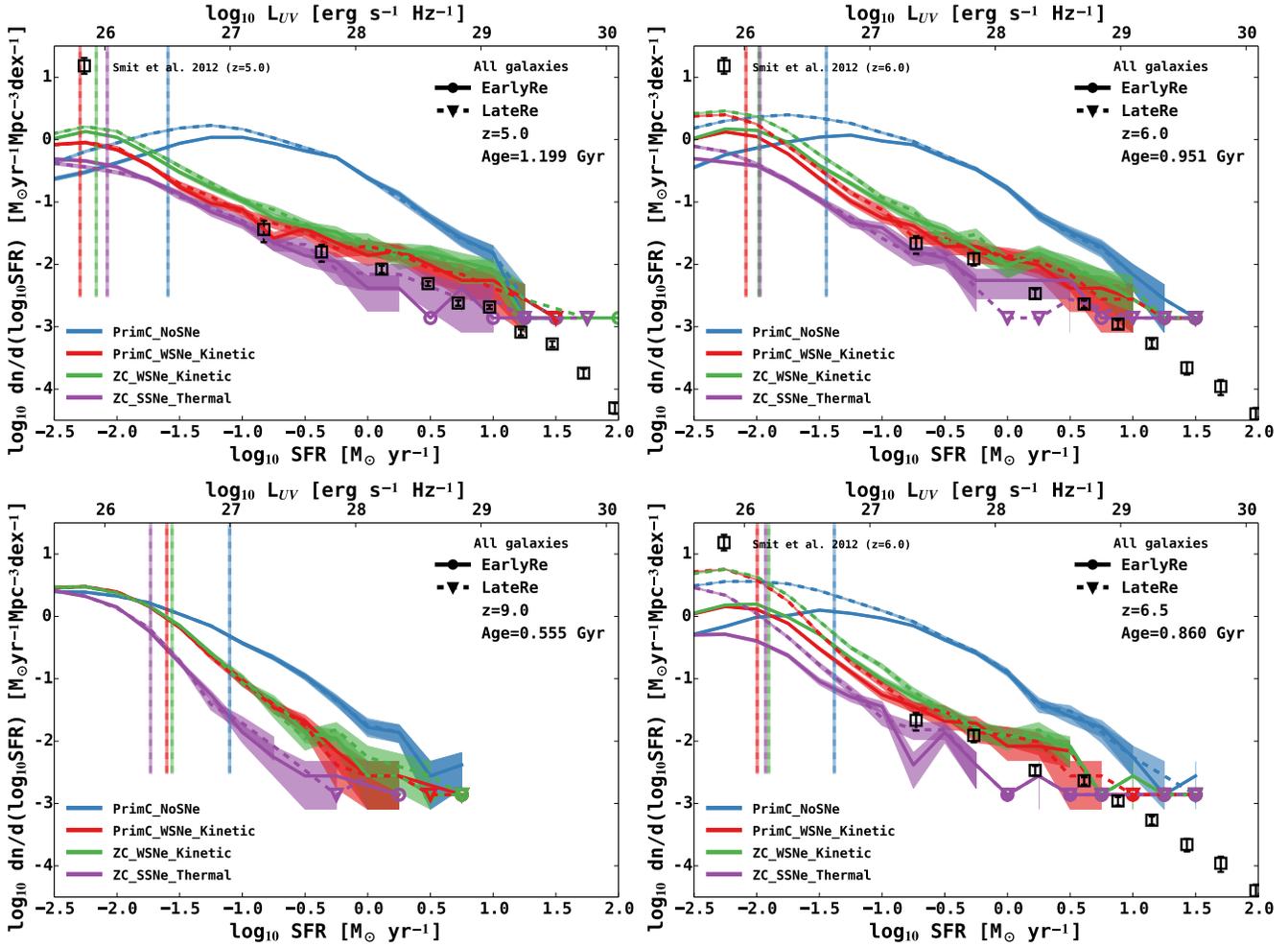

\centering
\begin{subfigure}{0.495\textwidth}
\includegraphics[width=\textwidth,keepaspectratio=true]{{{plots/SFRZ_newsmit_0512_z5.000002_All}}}
\end{subfigure}
\begin{subfigure}{0.495\textwidth}
\includegraphics[width=\textwidth,keepaspectratio=true]{{{plots/SFRZ_newsmit_0512_z6.006774_All}}}
\end{subfigure}

\begin{subfigure}{0.495\textwidth}
\includegraphics[width=\textwidth,keepaspectratio=true]{{{plots/SFRZ_newsmit_0512_z9.026387_All}}}
\end{subfigure}
\begin{subfigure}{0.495\textwidth}
\includegraphics[width=\textwidth,keepaspectratio=true]{{{plots/SFRZ_newsmit_0512_z6.489635_All}}}
\end{subfigure}
\caption{The SFRD at $z{\sim}5.0,\, 6.0,\, 6.5,\, 9.0$ (top left clockwise) for different simulations, errors are Poissonian and indicated by the faint band around each of the curves. Typically for galaxies with high (${>}10 \Msol \rm {yr}^{-1}$) SFR, there is only one object in the binning range which we represent with a circle (down triangle) for early (late) reionization models. The galaxies at the very highest (${>}30 \Msol \rm {yr}^{-1}$) SFR are mergers. The solid (dashed) curves are the models with an instantaneous reionization at $\zr{=}9$ ($6.5$), while the colours represent the different simulation physics runs; \emph{PrimC} meaning primordial cooling only, \emph{ZC} means metal lines can contribute to the cooling rate too. \emph{NoSNe} is no feedback, \emph{WSNe} is the weak kinetic feedback, \emph{SSNe} is the strong thermal feedback. The shaded regions are $1\sigma$ confidence limits estimated by bootstrap.  The observations in open squares are from Smit (private communication) who applied the methodology of~\citet{Smit:12} with
dust-corrections of~\citet{Bouwens:13} to the latest rest-frame UV continuum observations of galaxies from~\citet{Bouwens:14}. We then corrected the SFR to a~\citet{Chabrier:03} IMF. 
The vertical lines are the approximate resolution limits of the simulations, below which the SFR will be increasingly underestimated.}\label{fig:SFR_tbox}.
\end{figure*}

\subsection{Implementation of reionization}
The effect of reionization is to suppress the contribution of faint end galaxies through photoevaporation of the gas as well as prevention of further accretion from the 
IGM~\citep[e.g.][]{Shapiro:04,Iliev:05}.
To test the impact of reionization in our simulations we ran two different sets of models. 
In one case we assumed an early ($\zr{=}9$) reionization (labelled \emph{EarlyRe}) and the
other a late ($\zr{=}6.5$) reionization (labelled \emph{LateRe}). These redshifts are reasonable 
bracketing cases from CMB and Lyman $\alpha$ forest bounds, e.g. Fig. 12 of~\citet{Pritchard:10}.
Following~\citet{Pawlik:09a} and~\citet{Schaye:10}, we model reionization as an instantaneous event
across the box, which is appropriate for the small volume of our simulation which would lie 
within a single \HII region during reionization~\citep{Furlanetto:04}. 

At reionization, all gas is exposed to a uniform~\citet{HaardtMadau:01} UV/X-ray background. 
We do not model radiative transfer in this work and instead calculate photoheating rates in the optically thin limit. 
At reionization, this will lead to an underestimate of the IGM temperature~\citep[e.g.][]{Abel:99} and we follow~\citet{Pawlik:09a} and~\citet{Wiersma:09a} by injecting 2eV per proton of additional thermal energy to ensure the gas is heated to $T{=}10^4\rm K$.
After reionization ($z{<}\zr$) all cooling rates are calculated in the presence of the evolving photoionizing~\citet{HaardtMadau:01} UV/X-ray background, as given in~\citet{Wiersma:09a}. 
For redshifts higher than $z{=}9$ the cooling rates are calculated in the presence of both the CMB and 
photodissociation from the $z{=}9$ background (as~\citealt{HaardtMadau:01} did not tabulate for $z{>}9$) with 
a cutoff in the spectrum above 1 Ryd. 
This `softer' background will suppress the formation of molecular hydrogen by the first stars/ionizing sources~\citep{Pawlik:09a}.

\section{Origin of UV Photons}
\label{sec:sfr}
In this section we consider the galaxy formation physics that is required to match 
the distribution of SFRs (the SFR function, Section~\ref{sec:sfrfn}), and the efficiency of star formation (i.e. sSFRs, Section~\ref{sec:ssfr}). 

\subsection{SFR function}\label{sec:sfrfn}
We first compare the SFR density (SFRD) function of the models with observational results 
from Smit (private communication) who applied the methodology of~\citet{Smit:12}, 
and the dust-corrections of~\citet{Bouwens:13}, to the latest 
rest-frame UV continuum observations of galaxies from~\citet{Bouwens:14} using
{\it HST} data in the HUDF09 and CANDELS 
fields~\citep{Grogin:11,Koekemoer:11,Bouwens:12a}. 
The quoted errors do not include the uncertainties due to imperfect dust-corrections and hence
discrepancies between simulation and observations are likely less severe than they appear. 
In Fig.~\ref{fig:SFR_tbox}, we show the observations and the simulation results for representative redshifts at 
$z{\sim}5,\,6,\,6.5,\,9$ (top left clockwise). 

Broadly speaking, the simulations that include feedback are in good agreement with observations below $10 \Msol \rm {yr}^{-1}$ at all redshifts (a similar result 
at larger mass scales has also been seen in~\citealt{Tescari:14}).
The thermal SNe scheme (\emph{ZC\_SSNe\_Thermal}; purple curve) is able to match the SFRD function at $z{\sim}5-6$ (top panels of Fig.~\ref{fig:SFR_tbox}).
Although at $z{\sim}6.5$ this model slightly underproduces the number of objects with SFR ${\sim}1 \Msol \rm {yr}^{-1}$ from Smit (private communication).

At the very highest SFRs (${>}30 \Msol \rm {yr}^{-1}$), the abundance of objects appears to be strongly overestimated in the simulations irrespective of the galaxy formation physics considered. 
In practice, within each SFR bin there is only one galaxy, indicated with a point and no errors (Poissonian error would be unity), and the last two data points at $z=5$ for the very highest rates are a result of
an ongoing merger. In a small simulation, volume statistics are limited for the highest SFRs and they will suffer from cosmic variance, $\sigma_{\rm V}$. To estimate how significant these simulated outliers are from the observed SFR function we use~\citet{Somerville:04} to estimate $\sigma_{\rm V}{=}b\sigma_{\rm DM}$, where $b$ is bias of the population and $\sigma_{\rm DM}$ the variance of the underlying DM halo. The highest SFR of  $\sim 50 \Msol \rm {yr}^{-1}$ at $z=5$ is in a halo of mass $2.86\times 10^{11} \hMsol$, which is calculated~\citep{pynbody} to have a bias $b = 6.5$. From~\citet{Somerville:04} we have $\sigma_{\rm DM}{\sim}0.3$ for our boxsize, giving  $\sigma_{\rm V}{=}b\sigma_{\rm DM}{=}1.8$, which makes the final data points highly susceptible to cosmic variance and thus disagreements with the observations are of negligible statistical significance. Why the merger triggered starbursts are visible in the simulations so often is suggestive that we are lacking an additional suppression of star formation in the largest systems, perhaps through supermassive black hole feedback~\citep[e.g.][]{Booth:09}.
We will test the impact of supermassive black holes on the galaxy properties in future work but caution that in the limited cosmological volume of our simulations we wouldn't expect to find quasars of significant black hole mass (from~\citealt{McGreer:13} a single object of mass ${\sim}10^9 \Msol$ would be found in a $\rm Gpc$ volume).

Finally, we caution that the observations may be underestimating the bright end of the SFRD function due to imperfect dust 
extinction corrections which become ever more significant for galaxies with large SFR and/or stellar masses~\citep[e.g.][]{Meurer:99,Santini:14}. Indeed, just such a rise in the
abundance of bright end galaxies has been seen in~\citet{Bowler:14} using additional data from UltraVISTA~\citep{Bowler:12,McCracken:12}. However, as argued by~\citet{Wyithe:11},
the exponential decline in the bright end of the luminosity function at $z{>}6$  is expected to disappear for galaxies ${<}-22$ mag due to gravitational lensing, 
which produces an observed power law luminosity function at bright luminosities \citep{Pei:95}. Due to the difficulties in uncovering the underlying luminosity function, 
as well as the lack of the most massive galaxies in the limited volume we simulate, we defer this analysis to a future, larger volume hydrodynamic simulation.

The simulations show the effect of radiative suppression of the low-mass (and low-SFR) galaxies by reionization
(comparing the two bottom panels of Fig.~\ref{fig:SFR_tbox}, before and after the early reionization scheme has occurred). 
However, the effects of reionization become visible in objects with SFRs an order of magnitude below the observational results of Smit (private communication)
shown in the bottom right panel of Fig.~\ref{fig:SFR_tbox}.

\begin{figure*}
\includegraphics[width=\textwidth,keepaspectratio=true]{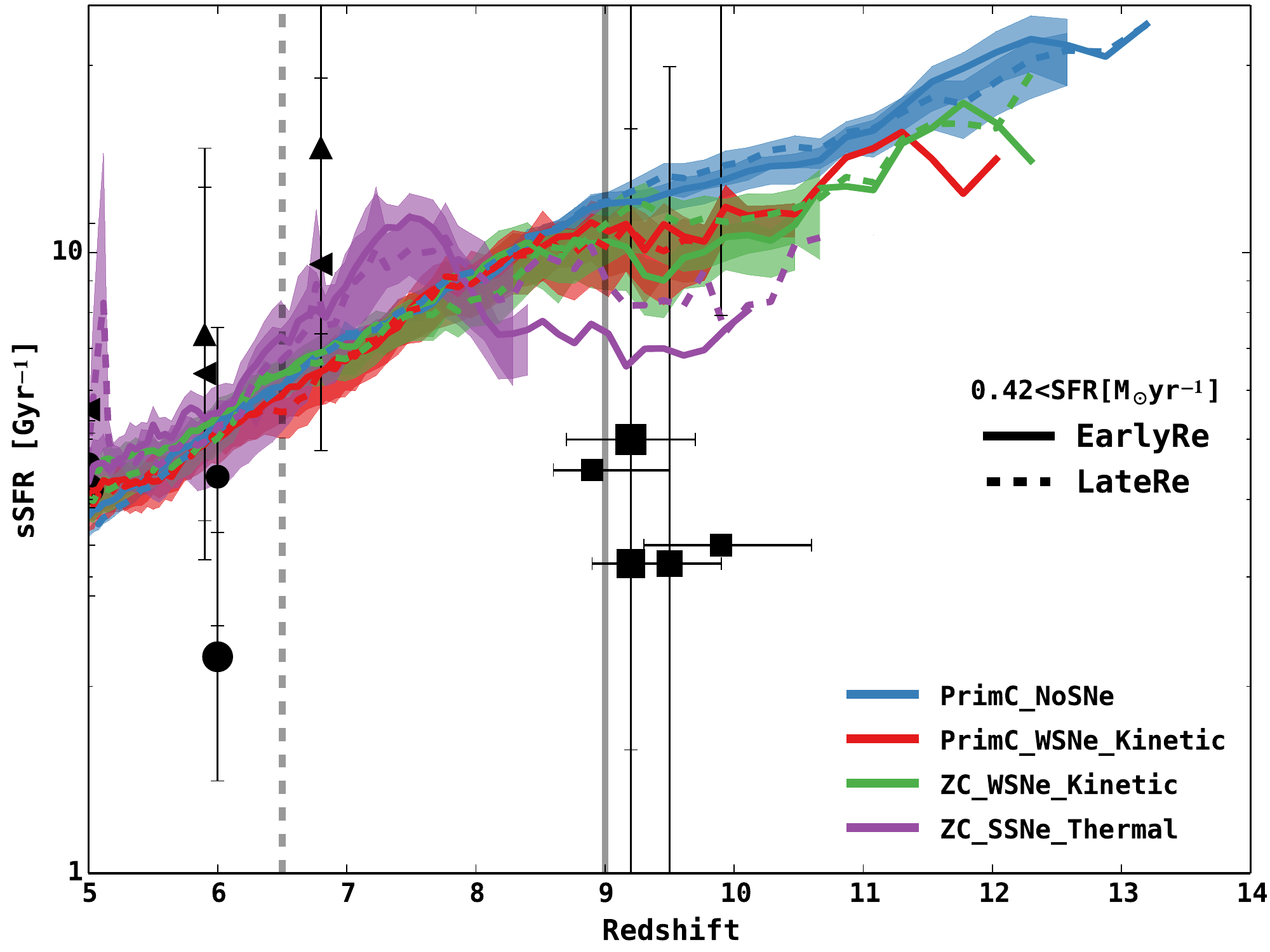} \caption{The sSFR rates as a function of redshift are now restricted for galaxies with an observationally derived cut of SFR $0.42 \Msol yr^{-1}$ (corrected to our Chabrier IMF from the quoted limit of $0.7$ using the Salpeter IMF). The solid (dashed) curves are the models with an instantaneous reionization at ${\zr}{=}9$ ($6.5$), while the colours represent the different simulation physics runs; \emph{PrimC} meaning primordial cooling only, \emph{ZC} means metal lines can contributed to the cooling rate too. \emph{NoSNe} is no feedback, \emph{WSNe} is the weak kinetic feedback, \emph{SSNe} is the strong thermal feedback. The shaded regions are $1\sigma$ confidence limits estimated by bootstrap.  The observations in black filled circles are from~\citet{Gonzalez:14} using a Constant SFH with nebulae emission line contaminations (and the point size represents the two stellar mass bins of $0.3$ dex, centred at $\log_{10} M_{*} \hMsol {\approx} 9.0$ and $9.7$); the black squares are the single galaxy observations from~\citet{Oesch:14} and~\citet{Holwerda:14}, the triangles are from~\citet{Stark:13} who corrected~\citet{Bouwens:12b} stellar masses for contamination by nebulae emission lines; the leftward (upward) pointing triangles are assuming a fixed (evolving) H$\alpha$ equivalent width. The main result is that, surprisingly, the {\it entire} simulation series match the observations. As discussed in the text, this is a simple result of the observational limit imposed on the simulations.}\label{fig:sSFR_sfrrange}
\end{figure*}

\begin{figure*}
\includegraphics[width=\textwidth,keepaspectratio=true]{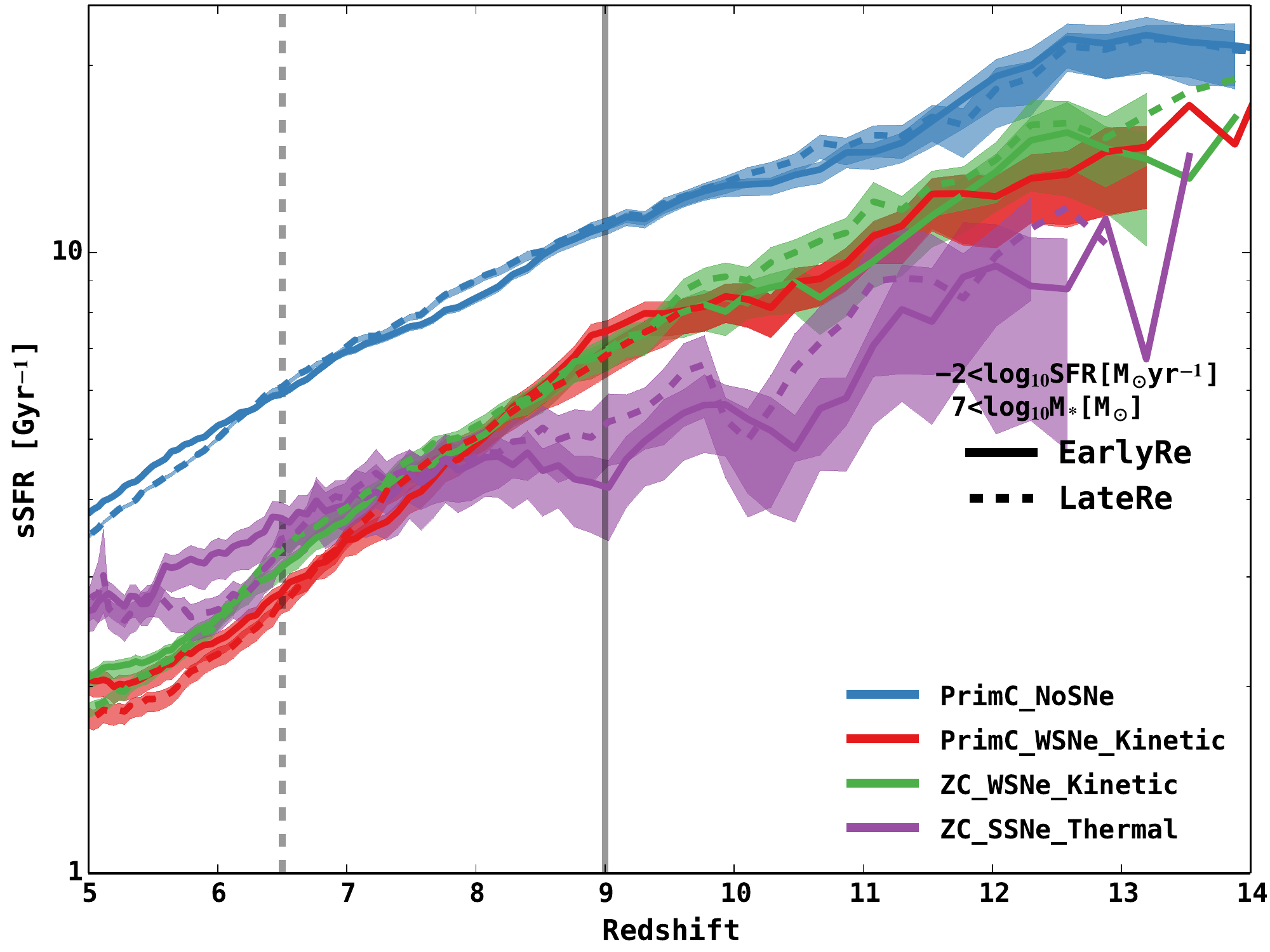} \caption{As in Fig.~\ref{fig:sSFR_sfrrange}, we show the sSFR as a function of redshift, for all objects above the resolution limit of the simulation as determined in Duffy et al. (in preparation) this is an SFR 40 times below the observational cut and a stellar mass at least two orders of magnitude below the limits of~\citet{Gonzalez:14}. 
The shaded regions are $1\sigma$ confidence limits estimated by bootstrap. We now see that the behaviour of the sSFR is a complex function of SNe feedback (sSFR is inversely proportional to increased feedback efficiency) , metal pollution of the IGM (resulting in enhanced cooling rates, and hence sSFR) and radiative feedback from reionization (typically a reduction of a factor of 5 in the sSFR). In particular, the timing of reionization results in a lasting impact on the sSFR, as described in the text. We have removed the observations
shown in Fig.~\ref{fig:sSFR_sfrrange} as they correspond to a different selection effect than the one considered in this figure. However, to
guide the relative change in the simulated sSFR we note that the blue  \emph{PrimC\_NoSNe} model is essentially equivalent in this figure and Fig.~\ref{fig:sSFR_sfrrange}.}\label{fig:sSFR_tbox_vlow}
\end{figure*}

\subsection{Efficiency of star formation}
\label{sec:ssfr}
The sSFR, along with the SFR, is a key variable observed at high-redshift. It is also an important prediction for  
infall-governed galaxy growth models which predict a rising sSFR with increasing redshift~\citep[e.g.][]{Bouche:10, Dave:11a}. 
In particular,~\citet{Haas:13a} found that the sSFR as a function of stellar mass was sensitive to the feedback model (their fig. 4, bottom-left panel) 
with a complex relation between sSFR and increasing stellar mass.

We emulate this test through first comparing the global sSFR with observations, and then extending the simulated sSFR measure to a minimum stellar mass.
We note that in the near future there is unlikely to be sufficient observations available to construct the full differential plot, i.e. sSFR versus $M_{\star}$, 
although future observations (utilizing gravitational lensing) may be possible.
Indeed, until recently most observations showed a rapid rise in the global sSFR from the present day to $z{\sim}2$ and then a plateau to higher redshifts~\citep[e.g.][]{Daddi:07,Noeske:07,Stark:09,Gonzalez:10}.
This picture has changed with the demonstration in~\citet{deBarros:14} that nebular line emission could contaminate the stellar mass estimates, and the most recent observations of~\citet{Stark:13} with this correction now demonstrate a continuing modest rise in sSFR for $z{>}4$. The observed sSFR is shown in Fig.~\ref{fig:sSFR_sfrrange} for galaxies with a minimum SFR of $0.42\Msol yr^{-1}$. The observational limits as given in~\citealt{Oesch:12b} are $0.7\Msol yr^{-1}$, which we reduce by a factor of 1.65 to account for the different stellar initial mass function assumed, as argued by~\citet{Haas:13a}. Note that we do not `correct' the sSFR as both SFR and stellar mass are reduced by the same factor and hence cancel out to first order.

For comparison, in Fig.~\ref{fig:sSFR_sfrrange} we show that the {\sc Smaug} simulations reproduce these observations, {\it irrespective of reionization history or feedback scheme}. 
We find that the sSFR in galaxies with SFR ${>}0.42\Msol yr^{-1}$ increases from $5\, [\rm Gyr^{-1}]$ at $z{=}5$ to ${\sim}10\, \rm Gyr^{-1}$ by $z{\sim}9$, 
in particular for the case of the no-feedback model (\emph{NoSNe} in blue) ultimately reaching ${\sim}20\, \rm Gyr^{-1}$ by $z{\sim}12$.
These values are in good agreement with a recent simulation by~\citet{Dayal:13}; however as we have tested
a range of physics schemes we see that this matching is irrespective of the galaxy formation physics modelled, 
and should not be considered a `success' of any one model. As discussed in~\citet{Dave:11b} for models without feedback (\emph{NoSNe} in blue), 
the only physics to impact the evolution of the sSFR is the accretion rate of material on to the growing haloes.
For those simulations with stellar feedback,~\citet{Dave:11b} 
argue that the insensitivity is due to the fact that the SN suppression of the SFR and growth in
the stellar mass is reduced by a similar factor, meaning that the ratio of the quantities is unchanged.
Therefore, we argue that this insensitivity (and the ability to match observations) is a direct result of the {\it observational limit} imposed on the simulated data.

\begin{figure*}
\includegraphics[width=\textwidth,keepaspectratio=true]{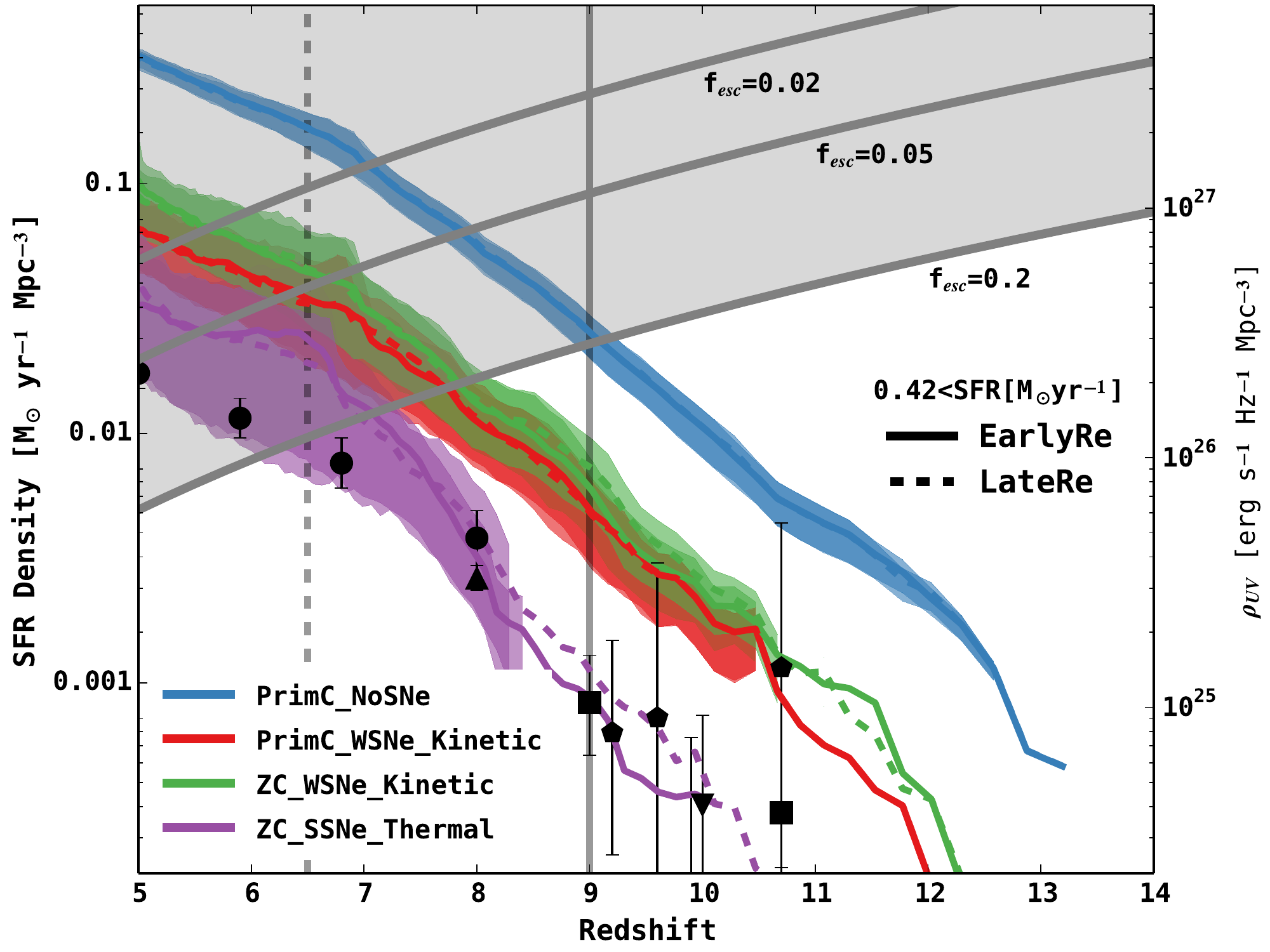} \caption{The Madau diagram (SFRD as a function of redshift) is presented with an observational cut of SFR $0.7\Msol yr^{-1}$ (converted to the Chabrier IMF this is $0.42$). The simulation which uses all of the available SNe energy (\emph{ZC\_SSNe\_Thermal}) to drive an outflow, together with the observationally inferred cut, is in excellent agreement with the observations. The solid (dashed) curves are the models with an instantaneous reionization at $\zr{=}9$ ($6.5$), while the colours represent the different simulation physics runs; \emph{PrimC} meaning primordial cooling only, \emph{ZC} means metal lines can contribute to the cooling rate too. \emph{NoSNe} is no feedback, \emph{WSNe} is the weak kinetic feedback, \emph{SSNe} is the strong thermal feedback. The shaded regions are $1\sigma$ confidence limits estimated by bootstrap. The observations in black filled circles are based on the HUDF~\citep{Bouwens:07, Bouwens:11b, Oesch:12b}, the squares are the HUDF and eXtreme Deep Field with GOODS-South~\citep{Oesch:13a}, the pentagon is from HUDF and eXtreme Deep Field with both GOODS-South {\it and} North~\citep{Oesch:14}, the upward pointing triangle is from BoRG~\citep{Bradley:12, Schmidt:14} and the downward pointing triangle is from CLASH~\citep{Bouwens:12a, Zheng:12, Coe:13}. The bottom grey horizontal curve indicates the UV ionisation density to completely reionize the Universe (hence case `A' recombination rates for hydrogen are used), assuming a clumping factor of 3~\citep{Pawlik:09a, Shull:12}, escape fraction of $0.2$~\citep{Ouchi:09} and a (logarithmic) ratio of Lyman continuum photons to UV flux of $25.2$~\citep{Robertson:13}. The other grey curves are for lower escape fractions.}\label{fig:Madau_sfrrange}
\end{figure*}

\begin{figure*}
\includegraphics[width=\textwidth,keepaspectratio=true]{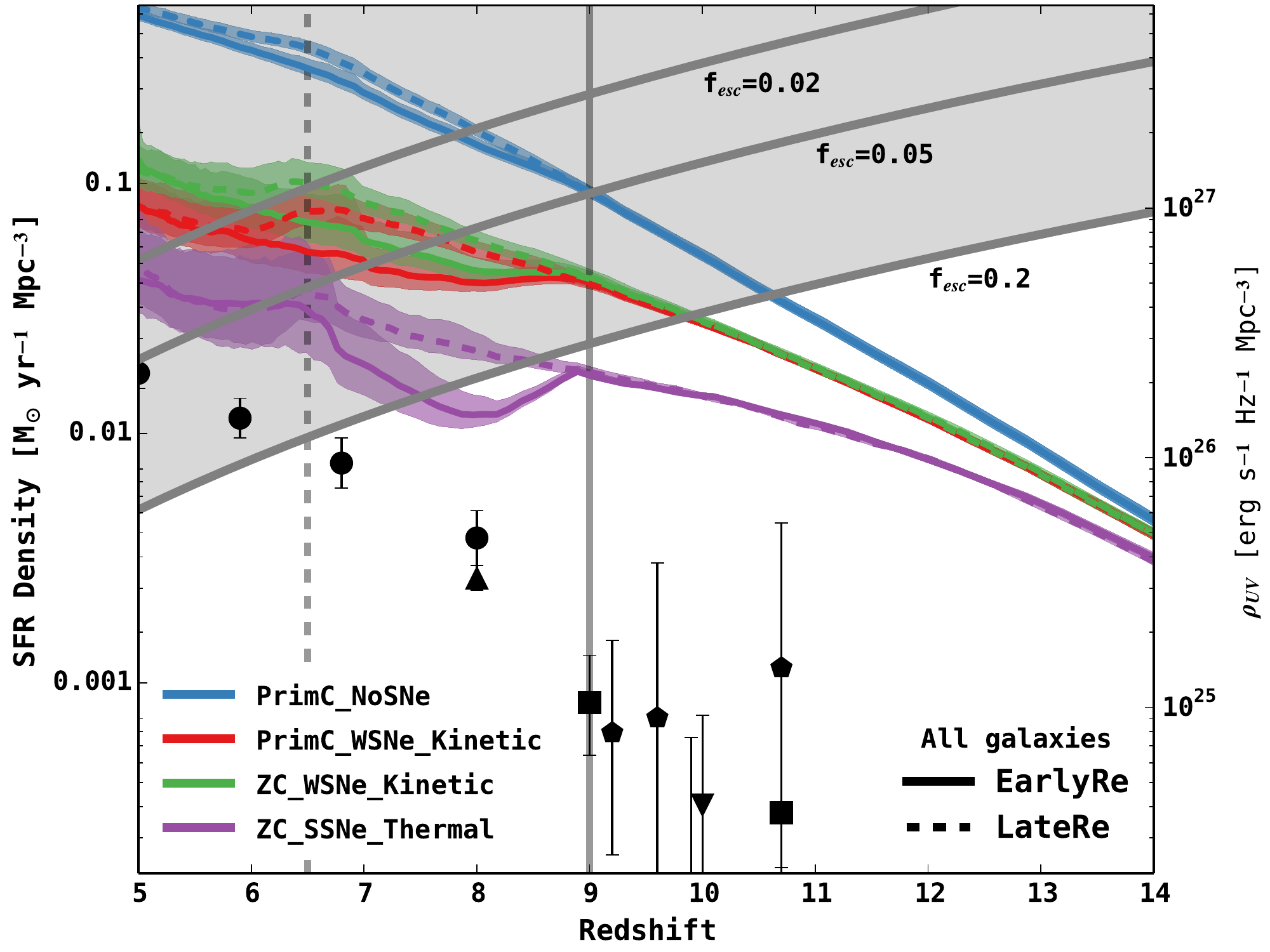} \caption{As with Fig~\ref{fig:Madau_sfrrange} we show the Madau diagram but now with the observationally motivated limit on SFR relaxed to include all galaxies in the simulation. The solid (dashed) curves are the models with an instantaneous reionization at $\zr{=}9$ ($6.5$), while the colours represent the different galaxy physics considered; \emph{PrimC} meaning primordial cooling only, \emph{ZC} means metal lines can contributed to the cooling rate too. \emph{NoSNe} is no feedback, \emph{WSNe} is the weak kinetic feedback, \emph{SSNe} is the strong thermal feedback. The shaded regions are $1\sigma$ confidence limits estimated by bootstrap.  The observations in black filled circles are based on the HUDF~\citep{Bouwens:07, Bouwens:11b, Oesch:12b}, the squares are the HUDF and eXtreme Deep Field with GOODS-South~\citep{Oesch:13a}, the pentagon is from HUDF and eXtreme Deep Field with both GOODS-South {\it and} North~\citep{Oesch:14}, the upward pointing triangle is from BoRG~\citep{Bradley:12, Schmidt:14} and the downward pointing triangle is from CLASH~\citep{Bouwens:12a, Zheng:12, Coe:13}. The grey horizontal curves indicates the UV ionisation density needed to completely reionize the Universe (hence case `A' recombination rates for hydrogen are used), assuming a clumping factor of 3~\citep{Pawlik:09a, Shull:12}, a plausible range of escape fraction~\citep{Wyithe:10} and a (logarithmic) ratio of Lyman continuum photons to UV flux of $25.2$~\citep{Robertson:13}.}\label{fig:Madau_tbox}
\end{figure*}

Fig.~\ref{fig:sSFR_tbox_vlow} shows the predicted sSFR in galaxies with SFR ${>}0.01$ and $M_{\star}{>}10^7\Msol yr^{-1}$.
These plots indicate that in a much deeper survey we would expect to be able to differentiate between feedback schemes, as they modify 
the typical star formation efficiency at lower galaxy stellar mass than currently observed.
The global results shown in Fig.~\ref{fig:sSFR_tbox_vlow} demonstrate three key behaviours:
\begin{itemize}
\item The overall sSFR is reduced as feedback efficiency is increased (from no-feedback, weak and ultimately strong SNe feedback; blue, red and purple curves respectively) in agreement with, for example,~\citet{Dave:11a}. At early times, $z=10$, there is a simple scaling between sSFR and SNe energy fraction $f_{\rm SNe}$ used in the simulation; $sSFR\approx 13/(1+f_{\rm SNe})\,{\rm Gyr^{-1}}$ where $f_{\rm SNe}$ is $0$ for \emph{NoSNe}, $0.4$ for \emph{WSNe} and \emph{SSNe} is $1$.
\item Pollution of the IGM by metals enhances the cooling rate, and results in a small ${\sim} 20\%$ increase in the sSFR~\citep[in qualitative agreement with][]{Haas:13a}. This increase occurs with a small delay (i.e. becoming apparent only below $z=8$) as it takes time for significant amounts of metals to be produced and released (comparing \emph{PrimC} and \emph{ZC} models; green and red curves respectively). 
\item There is a modest hysteresis in the sSFR as to {\it when} reionization occurred. The stellar mass in the denominator of sSFR is an integral of the SFR, meaning that the instantaneous SFR is reduced just after reionization, while the stellar mass is approximately the same as a run with a different reionization redshift. This effect is an $\sim 5-10\%$ reduction in the sSFR (comparing purple curves dashed and solid line of \emph{ZC\_SSNe\_Thermal} in Fig.~\ref{fig:sSFR_tbox_vlow}) which is still visible at the end of the simulation. This memory is in contrast with the global SFR alone which quickly converges after reionization and the models of different reionization redshifts converge by $z{\sim}5$ as shown in~\ref{fig:Madau_sfrrange} and Fig.~\ref{fig:Madau_tbox}).
\end{itemize}

\begin{figure*}
\includegraphics[width=\textwidth,keepaspectratio=true]{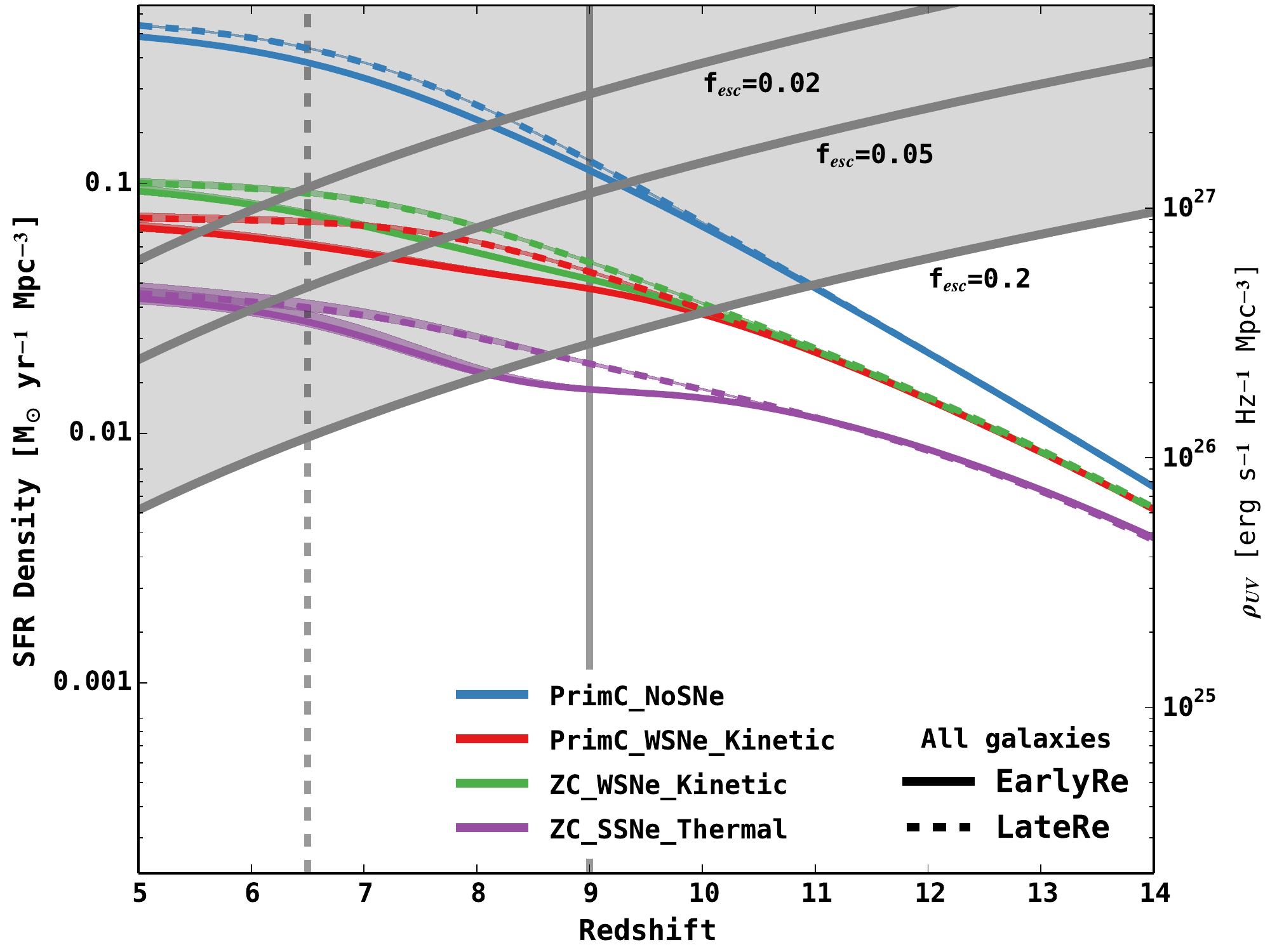} \caption{As with Fig.~\ref{fig:Madau_tbox} we show the Madau diagram with the observationally motivated limit on SFR relaxed to include all galaxies in the simulation, now focusing on the 
observability of the reionization `dip', and hence the redshift range has been restricted. 
The solid (dashed) curves are the models with an instantaneous reionization at $\zr{=}9$ ($6.5$), while the colours and represent the different galaxy physics considered; \emph{PrimC} meaning primordial cooling only, \emph{ZC} means metal lines can contribute to the cooling rate too. \emph{NoSNe} is no feedback, \emph{WSNe} is the weak kinetic feedback, \emph{SSNe} is the strong thermal feedback. The shaded regions are $1\sigma$ confidence limits estimated by bootstrap. The grey horizontal curves indicates the UV ionisation density needed to completely reionize the Universe (hence, case `A' recombination rates for hydrogen are used), assuming a clumping factor of 3~\citep{Pawlik:09a}, a plausible range of escape fraction~\citep{Wyithe:10} and a (logarithmic) ratio of Lyman continuum photons to UV flux of $25.2$~\citep{Robertson:13}. We have applied a Gaussian weighted average across the lines with a $\sigma$ of $\Delta z {\approx} 1$ to account for the variance in reionization due to the limited volume simulated. The sharp `dip' seen in Fig.~\ref{fig:Madau_tbox} is now a more gradual feature as the data is averaged over a larger redshift range. The error bars are also averaged in this measurement though and the `dip' between schemes with early and late reionization remains easily discernible. }\label{fig:Madau_dip}
\end{figure*}

\section{Sustaining reionization}
\label{sec:sfrd}
In Fig~\ref{fig:Madau_sfrrange}, we show the global SFRD as a function of redshift (also known as a Madau diagram) for our models.
To compare the simulation SFR with observations which measure UV luminosity, we have 
made use of the conversion between UV luminosity and SFR given by~\citet{Robertson:13}
\begin{equation}
\label{eqn:conv_luv_to_sfr}
L_{\rm UV} \approx 1.25 \times 10^{28} \times \frac{\rm{SFR}}{\Msol \rm {yr}^{-1}} {\rm erg\, s^{-1}\, Hz^{-1}}\,
\end{equation}
which is valid for a~\citet{Chabrier:03} IMF at solar metallicity\footnote{We adopt this 
solar metallicity value for ease of comparison with the assumptions in the literature, in particular~\citet{Robertson:13}, and we leave
the investigation of the mass-metallicity relation at these high-redshifts for a future paper.} (using a constant SFR~\citealt{BruzualCharlot:03} model).
In addition we incorporate the observational limits of inferred global measurements,
using a stated SFR limit of $0.7 \Msol yr^{-1}$ from~\citet{Oesch:12b}; they assume a~\cite{Salpeter:55} IMF which, when converted to a~\citet{Chabrier:03} IMF, corresponds to $0.42\Msol yr^{-1}$. 
In Fig.~\ref{fig:Madau_sfrrange} we see that all simulation schemes show a rising SFRD with time, but separate out depending on the level of SNe feedback.
In particular, our strongest SNe feedback model (\emph{ZC\_SSNe}; in purple) reproduces the observations at all redshifts.

For comparison, the grey curves indicate the level of SFRD (i.e. UV flux density) needed to sustain reionization at a given redshift, for a set of assumed escape fractions, $f_{\rm esc}$ (the fraction of ionizing
photons produced by stars that can escape their galaxy and participate in ionizing the IGM). 
We calculate this UV density by solving for the evolution of the volume filling fraction
of ionized hydrogen $Q_{\HII}$. This is a balance between a `source', the density of ionizing UV photons $\dot{n}_{\rm ion}$,
and 'sinks', recombinations in gas of density $\bar{n}_{\rm H}$ at a volume averaged recombination time $t_{\rm rec}$~\citep{Madau:98} 
\begin{equation}
\label{eqn:evolv_q}
\dot{ Q }_{\HII} = \frac{\dot{n}_{\rm ion}}{\bar{n}_{\rm H}} - \frac{Q_{\HII}}{\bar{t}_{\rm rec}} \,.
\end{equation}
At the completion of reionization, this equation simplifies to the steady state result of $\dot{ Q}_{\HII}{=}0$ and $Q_{\HII}{=}1$ such that the ionizing photon density needed to sustain reionization is
\begin{equation}
\label{eqn:steady_q}
\dot{n}_{\rm ion} = \frac{\bar{n}_{H}}{\bar{t}_{\rm rec}} \,.
\end{equation}
If we assume that the majority of ionizing photons come from massive stars in galaxies, then we can parameterise the original UV flux density needed to supply this density of ionizing photons
by assuming an escape fraction ($ f_{\rm esc} $) and a ratio of Lyman continuum to UV flux of $\Sigma_{\rm ion}$ (that is IMF dependent) to give $\rho_{\rm UV}{=}\dot{n}_{\rm ion} / (f_{\rm esc}\Sigma_{\rm ion})$. 
This results in
\begin{equation}
\label{eqn:luv_limit}
\rho_{\rm UV}  = \bar{n}_{\rm H} / (t_{\rm rec} f_{\rm esc} \Sigma_{\rm ion}) \,,
\end{equation}
where $\bar{n}_{\rm H}{=}X_{\rm p} \Omega_{\rm b} \rho_{\rm crit}$ is the mean hydrogen density, with $X_{\rm p}$ the primordial abundance fraction of hydrogen ${\sim}0.75$. The volume 
averaged recombination time $t_{\rm rec}$ is given by~\citep[e.g.][]{Kuhlen:12}
\begin{equation}
\label{eqn:trec}
t_{\rm rec} = \frac{1}{ C_{\HII} \alpha_{\rm recomb}(T) (1 + Y_{\rm P} / 4X_{\rm p}) \bar{n}_{\rm H} (1+z)^{3}} \,,
\end{equation}
where $C_{\HII}$ is the clumping factor, $Y_{\rm P}$ is the primordial abundance fraction of helium (effectively $1-X_{\rm p}$), 
$\alpha_{\rm recomb}(T)$ is the hydrogen recombination rate for an IGM temperature of $T{=}20,000K$~\citep[e.g.][]{Hui:03} 
for which we use case `A' (representing an entirely ionized IGM at the end of reionization).
We assume a clumping factor of 3~\citep{Pawlik:09a}, a (logarithmic) ratio of Lyman continuum photons to UV flux of $ \log_{10} \Sigma_{\rm ion}{=}25.2$~\citep{Robertson:13}
and escape fractions of $f_{\rm esc}{=}0.02$, $0.05$ and $0.2$ that span the likely range~\citep{Wyithe:10}. We note that our assumption of a clumping factor of 3 has been found by~\citet{Shull:12} to
well describe the global mean clumping factor over the redshift range $5{<}z{<}9$ which they find drives them to high escape fractions $\sim 0.2$ to sustain reionization at $z{\sim}7$, a point we will
find is in good agreement with our strong supernova feedback model  \emph{ZC\_SSNe}.
The grey curves in Fig.~\ref{fig:Madau_sfrrange} show the resulting UV density required to sustain reionization (equation~\ref{eqn:luv_limit}), or equivalently the required SFRD 
for our assumed~\cite{Chabrier:03} IMF (by scaling with equation~\ref{eqn:conv_luv_to_sfr}).
We see that for a large value of $f_{\rm esc}{=}0.2$, all simulations produce enough ionizing 
photons by $z{\sim}6$ to ionise the IGM using only observed galaxies, in agreement with~\citet{Robertson:13}. In addition,~\citet{Finkelstein:12} demonstrated a similar
result that observed populations could sustain reionization by $z{\sim}6$ for the combination $C_{\HII}/f_{\rm esc}=10$ which for our default clumping factor of 3 corresponds
to an escape fraction of $f_{\rm esc}{=}0.3$, of similarly large size as we conclude.
 
However, in addition to galaxies with SFRs above the observational limits, we can include the star formation 
from all galaxies in the simulation volume.
In this case (Fig.~\ref{fig:Madau_tbox}) we find that the observations would have missed a large contribution of ionizing 
photons from fainter objects 
(an extrapolation of this form was assumed in~\citealt{Robertson:13}).
Indeed, our simulations suggest that most star formation at $z{\ge}5$ is below current observational limits (in agreement with~\citealt{Paardekooper:13}). 
In particular, we find that the model which accurately reproduced the observed 
Madau diagram (\emph{ZC\_SSNe}; purple) in Fig.~\ref{fig:Madau_sfrrange}, produces 
only 10\% of the total predicted global SFRD in galaxies above the observed limit. 
The majority of ionizing photons in this model are undetectable by current telescopes, 
yet it is this faint population that can achieve reionization by $z{\approx}8$ (Fig.~\ref{fig:Madau_tbox}). 

\subsection{Radiative feedback}\label{sec:radeffect}
We find that reionization, as modelled by a uniform UV/X-ray background, suppresses the total SFRD~\citep{Barkana:00} particularly in the case of strong SNe 
feedback \emph{ZC\_SSNe}, indicating that strong stellar feedback enhances reionization feedback in agreement with~\citet{Pawlik:09b}.
However, by $z{\sim}7.5$ even the \emph{ZC\_SSNe} simulation with a post-reionization~\citet{HaardtMadau:01} UV/X-ray background is able to produce enough ionizing photons to sustain reionization.

\begin{figure*}
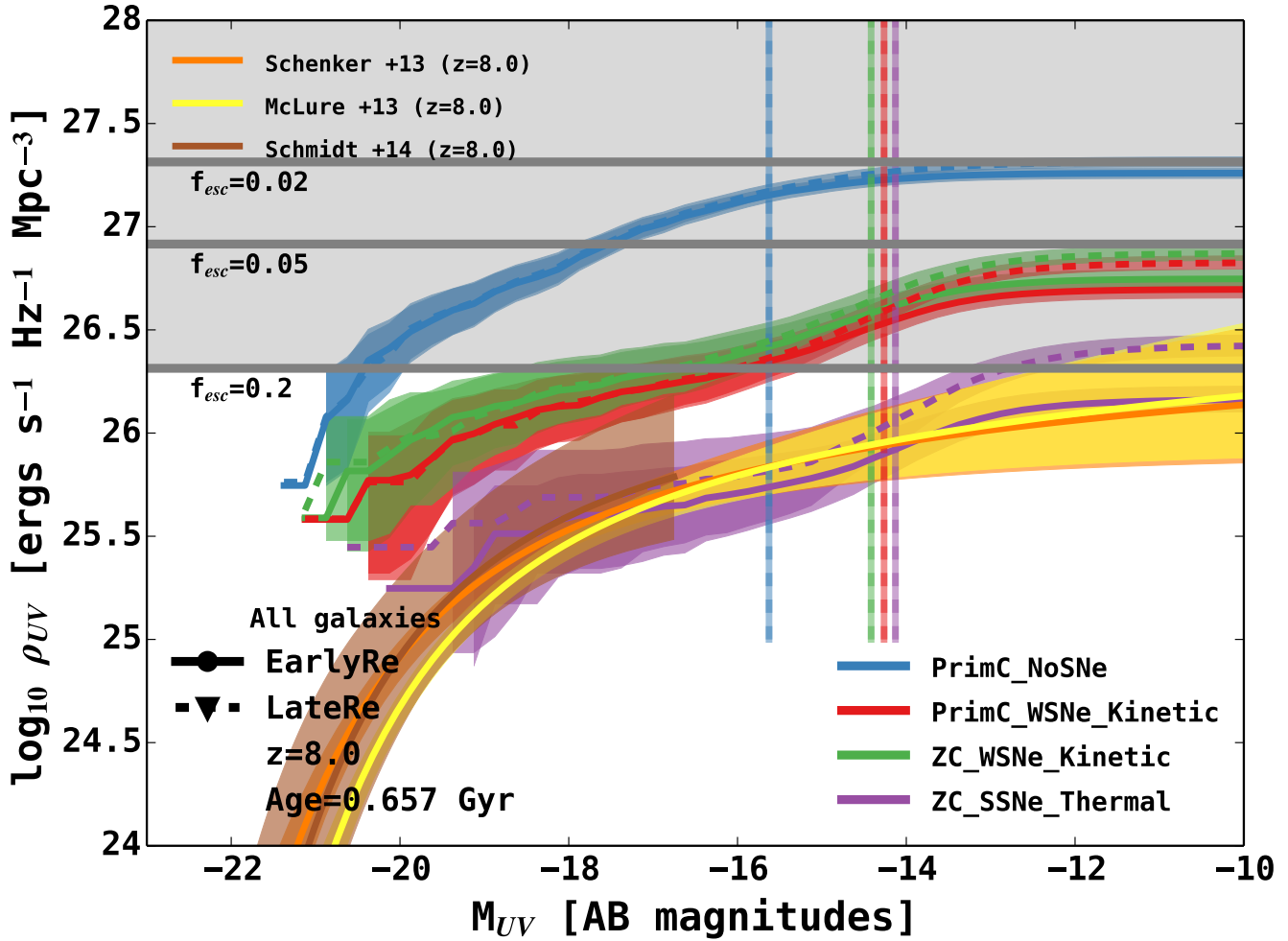

\includegraphics[width=\textwidth,keepaspectratio=true]{{{plots/ABMag_0512_z7.964357_All}}}
\caption{The cumulative UV luminosity density from galaxies of a given AB Magnitude at $z{\sim}8.0$ for different simulations, all of which can sustain reionization (the cumulative summation of their UV luminosity cross the grey curve, denoting the number of UV photons required for a given escape fraction). The solid (dashed) curves are the models with an instantaneous reionization at $\zr{=}9$ ($6.5$), while the colours represent the different simulation physics runs; \emph{PrimC} meaning primordial cooling only, \emph{ZC} means metal lines can contribute to the cooling rate too. \emph{NoSNe} is no feedback, \emph{WSNe} is the weak kinetic feedback, \emph{SSNe} is the strong thermal feedback. The shaded regions are $1\sigma$ confidence limits estimated by bootstrap. The observations in orange (yellow) bands are from~\citet{Robertson:13} who estimated the $1\sigma$ error region from the luminosity function of~\citet{Schenker:13} (\citealt{McLure:13}, respectively). The vertical lines are the approximate resolution limits of the simulations, below which the SFR will be increasingly underestimated.}\label{fig:ABMag_tbox}.
\end{figure*}

The progress of reionization does not occur instantaneously across the Universe at a given redshift as we have modelled. Instead, certain regions will reionize earlier than others, causing the global SFRD 
to be an average over different stages of reionization as larger volumes are considered. This means that the `dip' seen in one volume may be `smoothed' away. 
We can model this averaging by convolving the SFRD with a Gaussian function of width $\Delta z$ given by the prescription of~\citet{Barkana:04} and~\citet{Wyithe:04b} 
in which regions that are over-dense will be ionized before those that are under-dense. 

The $1\sigma$ cosmic variance in redshift at which a particular stage of structure formation is achieved amongst different 
simulations with finite volume was calculated by~\citet{Barkana:04}. From equation 2 (seen in Fig. 3) of~\citet{Barkana:04}, we find that for our volume of 
$(10\hMpc)^{3}$ there is a redshift variance of $\Delta z {\approx} 1$.
In Fig.~\ref{fig:Madau_dip}, we demonstrate the effect of this smoothing on the Madau diagram from Fig.~\ref{fig:Madau_tbox}.
We see that the sharpness of the dip is smoothed in practice if a larger volume was to be modelled, extending the effective duration of reionization. 
However, the error bars are also averaged (and hence reduced), therefore the `dip' remains easily discernible in this figure still.

\section{Extrapolation to Faint Galaxies}\label{sec:faintgal}
\citet{Robertson:13} recently extrapolated measurements of the UV luminosity functions from~\citet{Schenker:13} and~\citet{McLure:13} to fluxes below the observed limit of AB
magnitude $M_{\rm AB}{=}-18$~\citep{Ellis:13,Koekemoer:13} and found that there were indeed enough ionizing photons to sustain reionization if galaxies as 
faint as AB magnitude $M_{\rm AB}{=}-13$ were considered. A similar conclusion was found by~\citet{Schmidt:14} who extrapolated the BoRG (Brightest of Reionizing Galaxies) survey
data~\citep{Bradley:12} to $M_{\rm AB}{=}-15$. In this section we make use of the high-resolution of our simulations to test the validity of these extrapolations.

In Fig.~\ref{fig:ABMag_tbox}, we have summed the UV flux from galaxies as a function of decreasing AB magnitude, bootstrapping to calculate $68\%$ confidence intervals around each
of the median lines. Results are shown at $z{=}8$. The simulation scheme with strong SNe feedback (\emph{SSNe} in purple) is again 
in good agreement with the observations, tracking the observed luminosity functions of~\citet{Schenker:13} and~\citet{McLure:13} shown in gold and yellow respectively, 
to their detection threshold and thereafter closely following the extrapolated luminosity function of~\citet{Robertson:13}. We have also added the BoRG survey data from~\citet{Schmidt:14} as a brown curve, 
who focussed on a brighter sample of galaxies than the other works, confirming the excellent agreement of our strong supernova feedback model (\emph{ZC\_SSNe\_Thermal}) with observations at $z=8$.

The purple curve (\emph{ZC\_SSNe\_Thermal}) crosses above the grey line corresponding to an ionizing photon escape fraction 
of $0.2$ (within the likely range) indicating that reionization can be sustained at $z{\sim}8$ just above a luminosity of $-13$ AB magnitude.

The ability of a faint end extrapolation to sustain reionization is critically dependent on the assumed escape fraction. Instead of assuming an escape fraction of
20\% (as was done in the observational work of~\citealt{Robertson:13} for example) we are more conservative and choose $f_{\rm esc}=0.05$. In this case the 
strong SNe scheme \emph{ZC\_SSNe\_Thermal} does not produce enough photons to sustain reionization until $z{\la}6$ which is at the limit of the lowest
possible redshift set by the~\citet{Gunn:65} trough in higher redshift quasar spectra. As the \emph{ZC\_SSNe\_Thermal} model reproduces the observed SFR function,
this is suggestive that either the unobserved faint end is significantly steeper than we predict in this model, or the escape fraction has to be higher than $f_{\rm esc}=0.05$ 
(it could of course be mass dependent in which case the constraints on the overall escape fraction are weakened).

\subsection{Caveats}
As we are unable to model self-shielding, we are liable to overestimate the impact of reionization because gas will be heated
that otherwise would not be exposed to the external UV/X-ray background. 
In addition, at the lowest luminosities, we are subject to resolution limits.
In Fig.~\ref{fig:ABMag_tbox} we show with vertical lines the approximate resolution limits for each simulation (determined in Duffy et al. in preparation). 
These limits indicate the luminosity at which the simulations will progressively begin to underestimate the galaxy counts.
Therefore, because we are likely to progressively underestimate the ever fainter end below these limits, our simulations can only underestimate the flux from 
galaxies to $M_{\rm AB}{=}-13$, resulting in sufficient ionizing photons from this faint/unobserved population of galaxies to reionize the Universe by $z{\la}8$
($z{\la}6$) assuming an escape fraction of 20\% (5\%).

\section{Conclusion} 
\label{sec:conclusion}

We have presented a new simulation series called {\sc Smaug}, consisting of high-resolution hydrodynamical (SPH) and $N$-body simulations
with systematic variations of feedback, cooling and reionization redshifts. We have shown
that a strong SNe feedback scheme can match the observed evolution in the SFRD as a function of redshift, provided care is taken
to model the observational threshold. When the ionizing photons from smaller, unobserved galaxies are included we find that this same model can sustain
reionization by $z{\sim}8$, with a faint end luminosity function in good agreement with extrapolations from current observations 
(e.g.~\citealt{Robertson:13}). 

The radiative feedback from reionization results in a factor of 2 suppression in the global SFRD. This suppression
is in galaxies that are currently unobserved, and is particularly efficient in models with strong SNe feedback (a result in agreement with~\citealt{Pawlik:09b}). 

Finally, we have demonstrated that {\it all} simulations, irrespective of feedback or reionization model, can match the observed rising sSFR as a
function of redshift. In contrast, the underlying global sSFR as a function of redshift is significantly more sensitive to the exact galaxy formation physics 
implemented. Specifically, we find that the sSFR is lowered as a function of increasing feedback strength, with a modest truncation post-reionization and a small divergence of $\Delta z {\approx}1$ 
between models with delayed reionization. This offers a potential constraint of the redshift of reionization through deeper measurements of the faint galaxy population at $z{\sim}5$.

\section*{Acknowledgements}
ARD would like to thank Dan Stark, Pascal Oesch and Brant Robertson for helpfully providing their results and discussions, Claudio Dalla Vecchia for his SNe feedback models, 
Chris Power and Doug Potter for their efforts with initial conditions, Steven Finkelstein, Katie Mack, Kasper Schmidt and Mike Shull for stimulating discussions and comments 
and a general thanks to OWLS teammates for their efforts in creating a wonderful simulation series package.
JSBW acknowledges the support of an Australian Research Council Laureate Fellowship. 
This work was supported by the Flagship Allocation Scheme of the NCI National Facility at the ANU, generous allocations of time through the iVEC Partner Share and Australian Supercomputer Time Allocation Committee. 
We would like to thank the {\sc Python} developers of {\sc matplotlib}~\citep{Hunter:07}, {\sc CosmoloPy} (http://roban.github.com/CosmoloPy) and {\sc pynbody}~\citep{pynbody} hosted at https://github.com/pynbody/pynbody for easing the visualization and analysis efforts in this work.

%\bibliographystyle{mn2e_warrick}
%\bibliography{full_references}

\label{lastpage}
\end{document}